\documentclass[aps, prd, twocolumn, superscriptaddress, nofootinbib]{revtex4-1}
\usepackage{graphicx}
\usepackage{dcolumn}
\usepackage{amssymb}
\usepackage{amsmath,amssymb,amsfonts}

\begin{document}

\newcommand{\Eq}[1]{\mbox{Eq. (\ref{eqn:#1})}}
\newcommand{\Fig}[1]{\mbox{Fig. \ref{fig:#1}}}
\newcommand{\Sec}[1]{\mbox{Sec. \ref{sec:#1}}}

\newcommand{\PHI}{\phi}
\newcommand{\PhiN}{\Phi^{\mathrm{N}}}
\newcommand{\vect}[1]{\mathbf{#1}}
\newcommand{\Del}{\nabla}
\newcommand{\unit}[1]{\;\mathrm{#1}}
\newcommand{\x}{\vect{x}}
\newcommand{\y}{\vect{y}}
\newcommand{\p}{\vect{p}}
\newcommand{\ScS}{\scriptstyle}
\newcommand{\ScScS}{\scriptscriptstyle}
\newcommand{\xplus}[1]{\vect{x}\!\ScScS{+}\!\ScS\vect{#1}}
\newcommand{\xminus}[1]{\vect{x}\!\ScScS{-}\!\ScS\vect{#1}}
\newcommand{\diff}{\mathrm{d}}
\newcommand{\mk}{{\mathbf k}}
\newcommand{\ep}{\epsilon}

\newcommand{\be}{\begin{equation}}
\newcommand{\ee}{\end{equation}}
\newcommand{\bea}{\begin{eqnarray}}
\newcommand{\eea}{\end{eqnarray}}
\newcommand{\vu}{{\mathbf u}}
\newcommand{\ve}{{\mathbf e}}
\newcommand{\vn}{{\mathbf n}}
\newcommand{\vk}{{\mathbf k}}
\newcommand{\vp}{{\mathbf p}}
\newcommand{\vx}{{\mathbf x}}
\newcommand{\PP}{{\mathbb P}}
\def\dup{\;\raise1.0pt\hbox{$'$}\hskip-6pt\partial\;}
\def\ddn{\;\overline{\raise1.0pt\hbox{$'$}\hskip-6pt\partial}\;}

\newcommand{\addressImperial}{Theoretical Physics, Blackett Laboratory, Imperial College, London, SW7 2BZ, United Kingdom}
\newcommand{\addressRoma}{Dipartimento di Fisica, Universit\`a di Roma ``La Sapienza'', P.le A. Moro 2, 00185 Roma, Italy}
\newcommand{\addressINFN}{Sez. Roma1 INFN, P.le A. Moro 2, 00185 Roma, Italy}
\newcommand{\addressRadboud}{Radboud University, Institute for Mathematics, Astrophysics and Particle Physics, Heyendaalseweg 135, NL-6525 AJ Nijmegen, The Netherlands}


\title{Parity at the Planck scale}
\author{Michele Arzano}
\affiliation{\addressRoma}
\affiliation{\addressINFN}

\author{Giulia Gubitosi}
\affiliation{\addressRadboud}
\affiliation{\addressRoma}

\author{Jo\~{a}o Magueijo}
\affiliation{\addressImperial}

\date{\today}

\begin{abstract}
We explore the possibility that well known properties of the parity operator, such as its idempotency and unitarity, might break down at the Planck scale. Parity might then do more than just swap right and left polarized states and reverse the sign of spatial momentum $\vk$: it might generate superpositions of right and left handed states, as well as mix momenta of different magnitudes. We lay down the general formalism, but also consider the concrete case of the Planck scale kinematics governed by $\kappa$-Poincar\'e symmetries, where some of the general features highlighted appear explicitly. We explore some of the observational implications for cosmological fluctuations. Different power spectra for right handed and left handed tensor modes might actually be a manifestation of  deformed parity symmetry at the Planck scale. Moreover, scale-invariance and parity symmetry appear deeply interconnected. 
\end{abstract}


\maketitle

\section{Introduction}

The search for the theory of quantum gravity has left us with the distinct possibility that familiar concepts, such as space-time manifolds, might completely dismantle at the Planck scale and be replaced by radically new structures (examples include non-commutative geometry, spin networks and multifractal theories \cite{Markopoulou:1997wi, Perez:2003vx, Majid:1994cy, Seiberg:1999vs, Ambjorn:2004qm, Ambjorn:2010ce, Calcagni:2009kc}). This prospect should serve as a warning against importing intuition derived from low-energy concepts into the UV/short-scale description of space-time. 
A case in point is the concept of parity at the Planck scale, and the fate of some of its familiar and ``self-evident'' properties associated with conventional, low-energy space-time.

In elementary treatments, the parity operator $\PP$ is frequently introduced as a transformation driven by a ``mirror'' action applied to the spatial reference frame, sending $\vx$ to $-\vx$, and then observing the transformation laws of all quantities, classical or quantum, that live in that space. But in most quantum gravity treatments the position space picture is heavily eroded, in some cases the arena of physics shifting in the first instance to a non-trivial momentum space (often curved), from which it is difficult, if not outright impossible, to derive a position space counterpart. Parity can then never be so simply associated with transformations in position space. Should the mathematical treatment point us in that direction, 
we should therefore not be afraid to eschew properties which are self-evident within the prejudiced
intuitions associated with position space.

In this paper we take aim at two such dogmas. Firstly, we challenge the idea that ${\mathbb P }^2=1$ is a logical necessity. That parity must be idempotent is obvious from the fact that two mirror transformations upon $\vx$ take us back to the original frame.
But once we abandon parity as a concept driven by a conventional pre-fixed smooth space, this need not be the case. In fact we find that lost idempotency of $\PP$ is a feature of models with $\kappa$-Poincar\'e relativistic symmetries. 

Secondly, we question whether parity merely swaps right-handed and left-handed particle states, whilst reversing their momentum. This property usually results from combining the idempotency and the unitarity (and thus hermiticity) of parity. Both could break down at the Planck scale. In either case parity would then map a right-handed state into a quantum superposition of right- and left-handed states. It might also bring the momentum $\vk$ non-trivially into the operation, and map a $\vk$ into a momentum of different modulus. 

In either case, the observational implications could be dramatic, should we have direct access to Planck scale physics, at least via  thought experiments. More mundanely, we could see parity at the Planck scale transmuted into parity violating tensor fluctuations left over from the early universe, as we show in this paper. 

The plan of this paper is as follows. In Section~\ref{general} we lay down the general framework for what might be the effects of parity at the Planck scale, and the most direct observational implications. 
For the rest of the paper we then illustrate {\it some} of the properties encoded into this general framework with reference to concrete theories of quantum gravity, or associated. Specifically, in Section III, we briefly introduce the $\kappa$-Poincar\'e algebra, a deformation of the ordinary algebra of relativistic symmetries that models putative quantum gravity effects at the Planck scale. We show how parity is non-trivially affected by the deformation of the algebra leading, as mentioned above, to an action of the parity operator on translation generators that no longer squares to one. We then explore the effects of the deformation on the helicity operator and show how parity swaps right handed and left handed states 
changing their spatial momentum. This leads to an interesting connection between parity invariance and scale invariance which we explore at the end of Section III. A discussion of the results is presented in the concluding Section IV.

\section{General framework and 
its phenomenology}\label{general}
We start by setting up the general framework for what could be the effects of parity at the Planck scale, should some of its basic properties be lost. We will at first do this
without reference to any specific theory (even though we mention possible sources for the effects), keeping the discussion as general as possible. 

It is usually the case that $\PP^2=1$, so that hermiticity and unitarity are equivalent: $\PP^\dagger=\PP=\PP^{-1}$. Given that $\PP^2=1$, its action on 2-dimensional vectors (such as the eigenvectors of helicity or of parity itself) can be encoded by a generic 
square root of the unit matrix (there are 4 such solutions):
\be\label{sqrtmat1}
 \alpha_{ij}=
  \left[ {\begin{array}{cc}
   a & \frac{1-a^2}{b} \\
   b & -a \\
  \end{array} } \right],
\ee
where $a$ and $b$ are any complex numbers, to begin with. Then, hermiticity forces $a$ to be real and 
$a^2 + |b|^2=1$, so that:
\be
 \alpha_{ij}=
  \left[ {\begin{array}{cc}
   a & b^\star \\
   b & -a \\
  \end{array} } \right]\quad , \quad a^2 + |b|^2=1.
\ee
If we are in the eigenbasis of parity itself then $b=0$ and 
$a$ can be set to $1$. If we are in the eigenbasis of  the helicity, then $a=0$ and $b$ can be at most a phase. Usually one sets $b=1$, so that
\be
 \alpha_{ij}=
  \left[ {\begin{array}{cc}
   0 & 1 \\
   1 & 0 \\
  \end{array} } \right].
\ee
The action of parity on a state with right/left (R/L) helicity therefore can only result in a state with opposite helicity, that is, parity merely swaps states with R and L helicities.

The situation is entirely different should the hermiticity and/or unitarity of the parity operator be lost (for example because they are defined with respect to a non-trivial measure \cite{AmelinoCamelia:1999pm}, so that with regards to the usual one they appear violated). This could happen:
\begin{itemize}
\item with  $\PP^2=1$ preserved, so that if we break one of hermiticity and unitarity we must break both (we {\it may} still preserve {\it both}, of course). 
\item  or with $\PP^2\neq 1$, in which case we are forced to break one of hermiticity or unitarity (or both, if we so wish). 
\end{itemize}
In either case parity can then map a R-helicity state into a superposition of R and L states, and likewise for L:
\bea\label{parRL}
\PP |R\rangle &=&\alpha_{RR} |R\rangle + \alpha_{RL} |L\rangle \\
\PP |L \rangle &=&\alpha_{LR} |R\rangle + \alpha_{LL} |L\rangle \,.
\eea
If $\PP^2\neq 1$ and parity is still unitary (but not hermitian) this is because there is no constraint on the rotation angle of the unitary operation. In this case parity invariance may become an infinite set of conditions, unless $\PP^n=1$ for some integer (in which case there are $n-1$ conditions). If $\PP^2=1$, but its hermiticity and unitary are lost, then parity is no longer an observable, so there is no point in seeking its eigenbasis. However helicity may still be an observable, in which case the action of parity upon the induced orthonormal basis is the most general matrix envisaged in Eq.(\ref{parRL}).

More generally, we should allow for the possibility that parity does not factor the internal space and the momentum $\vk$. Usually parity sends $\vk$ to $-\vk$ regardless of what it does to R and L states, and this could continue to the the case at the Planck scale, even if (\ref{parRL}) is non-trivial. But the action on momentum space could also be more complicated, and not factor it out of the action upon R and L. 
In general we should consider the matrix:
\be
\PP |i, \vk \rangle =\sum_{j,\vk '}\alpha_{ij}(\vk,\vk')  |j, \vk '\rangle,
\ee
with $i,j=R/L$.
The usual action of parity corresponds to:
\be
\alpha_{ij}(\vk,\vk') =\delta(\vk+\vk') 
  \left[ {\begin{array}{cc}
   0 & 1 \\
   1 & 0 \\
  \end{array} } \right]\,,
\ee
but this need not be true at the Planck scale. 

The different action of parity upon states would have immediate phenomenological implications, should the fluctuations of our Universe have their origin in quantum vacuum fluctuations.  
The vacuum expectation value of a given field may be seen as the norm of its one particle states 
(see \cite{Arzano:2015gda} and references therein):
\be
\langle \vk, R| \vk', R\rangle=\delta (\vk-\vk'){\cal P}_R(k)
\ee
and likewise for L. The way parity invariance usually forces ${\cal P}_R(k)={\cal P}_L(k)$ is by the following chain:
\bea\label{chain}
\langle \vk, R| \vk', R\rangle=\PP\left(\langle  \vk, R| \vk', R\rangle\right)=
\langle -\vk, L| -\vk' L\rangle
\eea
after which isotropy leads to ${\cal P}_R(k)={\cal P}_L(k)$ (since $|\vk|=|-\vk|$). This 
argument breaks down should parity act non-trivially, as we shall now see. We separate two extreme cases, one involving purely the internal space, the other purely momentum space. In general these two cases  could appear combined and even interact non-trivially.

\subsection{Tensor modes L/R asymmetry}
Let us suppose that $\PP^2=1$ is preserved, but that unitarity and hermiticity are lost. This could happen
because the inner product is defined by a deformed measure (resulting 
in apparent non-unitarity and non-hermiticity with respect to the standard measure), but any other explanations for non-unitarity and non-hermiticity, while preserving idempotency, will fall under the remit of this discussion. Then, any square root of the unit matrix is a candidate for parity, for example (\ref{sqrtmat1})
with $a$ and $b$ {\it any} complex numbers. 
Adjusting the last step of the argument (\ref{chain}) and using isotropy would then lead to:
\be
{\cal P}_R(k)={\cal P}_L(k)\frac{|\alpha_{RL}|^2}{1-|\alpha_{RR}|^2} \label{chiral}
\ee
that is, in general  a chiral asymmetric spectrum of tensor fluctuations. 
Such asymmetry would leave a mark in the TB and EB components of the CMB polarization, as described in~\cite{leecj}. The point made here is that such asymmetry could be nothing but precisely an {\it expression} of parity invariance, when hermiticity and non-unitary are lost or deformed. Since $\PP^2=1$ invariance under parity still results in a single condition. 

\subsection{Antipodean implications} \label{scaleinvariance}
Even if there is no non-trivial action on the internal space, there could be effects in momentum space. 
This could affect scalar as well as tensor fluctuations (allowing us to drop the R/L labels in what follows). In that case, 
a parity transformation usually maps $\vk$ into $-\vk$, so that it adds nothing to isotropy. However, as we will see explicitly in the next Section, in theories where relativistic symmetries are modified the 
law of addition of momenta is deformed and is generally written as $\vk\oplus \vk'$. This in turn implies that the concept of inverse momentum $-\vk$ is generalized to the antipode  $S( \vk)$, such that $\vk \oplus S(\vk)=0$. Because the deformed addition rule is nonlinear in the momenta, also $S(\vk)$  is in general a nonlinear function of $\vk$. Then parity might act as:
\be
\alpha_{ij}(\vk,\vk') =\delta(\vk\oplus\vk') 
  \left[ {\begin{array}{cc}
   0 & 1 \\
   1 & 0 \\
  \end{array} } \right]\,.
\ee
and clearly $\PP^2=1$ is lost. 

Consider now scalar fluctuations subject to invariance under the action of such a parity operator. 
In this case  parity invariance implies that: 
\be 
{\cal P}(k)= {\cal P}(|S(\vk)| )\,,
\ee
so that scale-invariance and parity invariance are equivalent for such theories. 
Then the theory is parity invariant iff $n_S=1$, i.e. if its fluctuations are scale-invariant (realized if the UV asymptotic dimension is 2 \cite{Arzano:2015gda}). This also resolves the issue of the effects of iterating $\PP$, if $\PP^2=1$ is lost. By iterating $\PP$ we get a set of conditions which are all equivalent to scale-invariance.

\section{Explicit examples from the $\kappa$-Poincar\'e algebra}

We now show how some of the features described above actually emerge in a concrete model of unconventional kinematics at the Planck scale: the $\kappa$-Poincar\'e algebra  \cite{Lukierski:1991pn, Lukierski:1992dt, Lukierski:1993wxa, Majid:1994cy}. In this scenario we deal with a deformation of ordinary relativistic symmetries based on a momentum space given by a submanifold of de Sitter space 
\be
-p_{0}^{2}+p_{i}^{2}+p_{4}^{2}=\kappa^{2}\,,\,\,\,\,\, p_{0}+p_{4}>0 \label{eq:constraint1}\,,
\ee
where $\kappa>0$ is a Planckian energy scale (see e.g. \cite{KowalskiGlikman:2004tz, Arzano:2010kz}). A possible realization of deformed kinematics is in terms of energy and spatial momentum given by the embedding coordinates above, $p_{\mu} = (p_{0},\vp)$. In this scenario symmetry generators obey the ordinary Poincar\'e algebra
\bea
[N_{i},P_{j}]&=& i\delta_{ij}P_{0}\,,\;\;[N_{i},P_{0}]=iP_{i}\\
\;[J_{i},J_{j}]&=&i\epsilon_{ijk}J_{k}\,,\;\;[J_{i},N_{j}]=i\epsilon_{ijk} N_{k}\\
\;[N_{i},N_{j}]&=& -i\epsilon_{ijk}J_{k}\,,
\eea
where $P_{0}$ and $P_i$ are generators of translations associated with {\it embedding} momenta $p_{\mu}$. The Casimir operators constructed from $P_{\mu}$ are just the usual ones $\mathcal{C}_1 = P^{\mu} P_{\mu}$ and $\mathcal{C}_2 = W^{\mu} W_{\mu}$ with $W_{\mu}$ the Pauli-Lubanski vector. Let us notice that other choices of translation generators associated to different sets of coordinates on de Sitter momentum space are possible and these in general will lead to an algebra with non-linear commutators and deformed Casimirs, as we will discuss below. Additional non-trivial features associated to $\kappa$-deformation  are due to the non-abelian group structure of momentum space (see \cite{Arzano:2014cya} for details). These reflect in the way quantum numbers associated to symmetry generators combine. In particular the momentum composition rules are now {\it non-abelian} 
\bea
(p\oplus q)_{0}&=& \frac{1}{\kappa}p_{0} (q_{0}+q_{4})+\kappa \frac{q_{0}}{p_{0}+p_{4}}+\frac{\vec p\cdot \vec q}{p_{0}+p_{4}}\\
(p\oplus q)_{i}&=&  \frac{1}{\kappa}p_{i} (q_{0}+q_{4})+q_{i}\,.
\eea
Such rules dictate also a non-trivial inversion for translation generators (and associated four-momenta) which would normally be realized in terms of a map $P_{\mu}\rightarrow -P_{\mu}$. Such inversion is now replaced by an antipodal map or {\it antipode} 
\bea
S(P)_{0}&=&-P_{0}+\frac{\vec P^{2}}{P_{0}+P_{4}}=-P_{4}+\frac{\kappa^{2}}{P_{0}+P_{4}} \label{antip1}\\
S(P)_{i}&=&-\frac{\kappa P_{i}}{P_{0}+P_{4}}\,.\label{antip2}
\eea
Algebraic consistency requires that also boost generators exhibit a non-trivial antipode
\be\label{antip3}
S(N)_{i}=-\frac{1}{\kappa}(P_{0}+P_{4})N_{i}+\frac{1}{\kappa}\epsilon_{ijk}P_{j}M_{k}\,,
\ee
while rotations are left untouched $S(M)_{i}=-M_{i}$. Notice that the non-trivial antipodes for translation generators square to one, $S(S(P)_{0})=P_{0}$, $S(S(P_{i}))=P_{i}$ while for boosts this is no longer true since one has $S(S(N))_{i}=N_{i}+\frac{3}{\kappa}S(P)_{i}\neq N_{i}$.

\subsection{The lost idempotency of parity in $\kappa$-Poincar\'e}
 Let us now turn to the problem of extending parity into the $\kappa$-deformed framework. In this context the intuition coming from the space-time picture of parity as space inversion is missing. Indeed the coordinate space counterpart of the structures we reviewed so far can be formulated in terms of a noncommutative space-time \cite{Majid:1994cy} leading to a much less intuitive picture and a rather evasive physical interpretation. Thus one has to rely mostly on the algebraic structure of symmetry generators and on the geometrical picture of momentum space. 
 
We here elaborate on two possible choices for the action of parity on symmetry generators. The most straightforward one would be simply to assume that parity is undeformed i.e. that its action is given by\footnote{Notice that we have refrained from writing the action of parity as $\mathbb{P}{P}_i \mathbb{P}^{-1}$ since this would tacitly imply that we are assuming an action as a unitary operator.}
\begin{align}
   \mathbb{P}({P}_i)  = -P_i, \quad \mathbb{P}(P_0) =P_0\nonumber\\
   \mathbb{P}({M}_i) = M_i, \quad \mathbb{P}(N_i) = -N_i\label{parity}\,.
\end{align}
One possible drawback of postulating the usual definition of parity on deformed translation generators is that now the total linear momentum of a particle and its parity mapped image would no longer be zero, leading to an evident asymmetry with the total spatial momentum of the system pointing towards the direction of the original particle:
\bea
(p\oplus \mathbb{P}( p))_{i}&=&  \frac{1}{\kappa}p_{i} (\mathbb{P}( p)_{0}+\mathbb{P}( p)_{4})+\mathbb{P}( p)_{i}\nonumber\\
&=&p_{i}\left(\frac{p_{0}+p_{4}}{\kappa} - 1\right)\,.
\eea
To avoid this  feature, the authors of \cite{Arzano:2016egk} proposed to make use of the notion of antipode for deformed translation generators, and thus define the action of parity as
\begin{align}
   \mathbb{P}({P}_i)  = S(P)_i, \quad \mathbb{P}(P_0) =-S (P)_0\nonumber\\
   \mathbb{P}({M}_i) =- S(M)_i, \quad \mathbb{P}(N_i) =S (N)_i\label{parityAK}\,,
\end{align}
with the antipodes given by \eqref{antip1}, \eqref{antip2} and \eqref{antip3} above.
The first question one could ask is whether the definitions above are compatible with the non-trivial geometry of momentum space and, in particular, if the action of parity does not bring us {\it out of the momentum manifold} described by equations \eqref{eq:constraint1}. This can be verified upon observing  that $P_{4}$ is strictly related to the particle's mass and so it is left unchanged by parity, $\mathbb{P}({P}_4) = {P}_4$, in both cases above. Thus we have that the standard action of parity \eqref{parity} complies with the constraints in \eqref{eq:constraint1}. For the deformed parity transformation \eqref{parityAK} we also have that both of the constraints in \eqref{eq:constraint1} are left invariant, for on-shell momenta. (Notice that this might in principle be a nontrivial issue also for charge conjugation defined as in ordinary field theory as $\mathbb{C}(P_0) =-P_0$).

While being compatible with the deformed momentum space structure, the deformed parity \eqref{parityAK} {\it does not} square to the identity. Indeed a simple computation shows that $\mathbb P(\mathbb P(P_{0}))\neq P_{0}$:
\begin{widetext}
\bea
\mathbb P(\mathbb P(P_{0})) &=& \mathbb P(-S(P_{0}))= 
-S(P_{0})-\frac{\sum_{i}S(P_{i})S(P_{i})}{-S(P_{0})+P_{4}} \nonumber\\
&=&
P_{0}-\frac{\vec P^{2}}{P_{0}+P_{4}}-\frac{\frac{\kappa^{2} \vec P^{2}}{(P_{0}+P_{4})^{2}}}{P_{0}-\frac{\vec P^{2}}{P_{0}+P_{4}}+P_{4}} 
=P_{0}-2 \frac{\vec P^{2}}{P_{0}+P_{4}}\left(\frac{  P_{0}^{2}+ P_{0}P_{4}-\vec P^{2}}{(P_{0}+P_{4})^{2}-\vec P^{2}}\right)
\eea
\end{widetext}
and $\mathbb P(\mathbb P(P_{i}))\neq P_{i}$:
\vskip-2in
\begin{widetext}
\bea
\mathbb P(\mathbb P(P_{i})) &=& \mathbb P(S(P_{i}))=-\frac{\kappa S(P_{i})}{-S(P_{0})+P_{4}}= -\frac{-\kappa^{2} \frac{ P_{i}}{P_{0}+P_{4}}}{P_{0}-\frac{\vec P^{2}}{P_{0}+P_{4}}+P_{4}} = \frac{\kappa^{2} P_{i}}{(P_{0}+P_{4})^{2}-\vec P^{2}}\,.
\eea
\end{widetext}
Thus we see that defining parity in terms of antipodal maps leads to the curious feature of {\it lost idempotency}: applying twice a parity transformation does not bring us back to the original one-particle state.

\subsection{The helicity operator}

Let us now look at the effect of parity on massless one-particle states, eigenstates of the helicity operator. Since on the four-momenta $p_{\mu}$ we have an action of the ordinary Poincar\'e algebra an helicity operator can be defined in a similar fashion as in standard relativistic quantum theory 
\be\label{hclassical}
\hat{h} = - \frac{\vec{P}\cdot \vec{J}}{|\vec{P}|}\,.
\ee
As in familiar QFT for photons one has that $\hat{h}^2 = 1$ and thus helicity eigenstates are characterized by eigenvalues $h=\pm1$. Photon helicity eigenstates correspond to circular polarization states $|L/R\rangle$. In usual relativistic kinematics the parity operator $\mathbb{P}$ switches the sign of $\hat{h}$ and thus it swaps $|L\rangle$ and $|R\rangle$ states.

It is easy to check that the ``undeformed" helicity operator above changes sign under {\it both} actions of parity \eqref{parity} and \eqref{parityAK}. Thus even in the presence of a non-trivial parity transformation $|L\rangle$ states are mapped into $|R\rangle$ states and viceversa, as in the usual theory. However, the action of the deformed parity  \eqref{parityAK} affects the energy of the states. In fact, for parity defined in terms of the antipode we have now that  
\be
\mathbb{P} |\vk, R\rangle = |S(\vk),L\rangle\,.
\ee

In terms of the notation used in section \ref{general}, this corresponds to having
\be
\alpha_{ij}(\vk, \vk')=\delta(\vk\oplus \vk')   \left[ {\begin{array}{cc}
   0 & 1 \\
   1 & 0 \\
  \end{array} } \right]\,.
\ee

\subsection{Parity and helicity in a different basis of the $\kappa$-Poincar\'e algebra}

All of the results above concerning the properties of parity and helicity in theories with $\kappa$-Poincar\'e symmetries were derived using the  embedding  momenta, defined by the relations \eqref{eq:constraint1}. Another popular choice is the so-called bicrossproduct basis, whose translation generators $\bar P_{\mu}$ are related to the ones of the embedding basis via
\bea
\bar P_{0} &=& \kappa \ln\left( \frac{P_{0}+P_{4}}{\kappa}\right)\\
\bar P_{i}&=& \kappa \frac{P_{1}}{P_{0}+P_{4}}\,. 
\eea
In this basis the symmetry generators obey a modified algebra
\bea
[N_{i},\bar P_{j}]&=& i\delta_{ij}\left( \frac{\kappa}{2}\left(1-e^{-2 \bar P_{0}/\kappa}\right)+\frac{|\vec{ \bar P}|^{2}}{2 \kappa} \right) -\frac{i}{\kappa} \bar P_{i} \bar P_{j}\nonumber\\
\;[J_{i},J_{j}]&=&i\epsilon_{ijk}J_{k}\,,\,\;\;\;\;[J_{i},N_{j}]=i\epsilon_{ijk} N_{k}\\
\;[N_{i},N_{j}]&=&-i\epsilon_{ijk}J_{k}\,,\;\;[N_{i},\bar P_{0}]=i\bar P_{i}\,,\nonumber
\eea
and translation generators have antipode:
 \bea
S(\bar P)_{0}&=&- \bar P_{0}\label{antipbic1}\\
S(\bar P)_{i}&=& -\bar P_{1}e^{\bar P_{0}/\kappa}\,.\label{antipbic2}
\eea

 It is interesting to re-work the results discussed in the previous subsections using the bicrossproduct basis.
 
The action of both the definitions of parity, \eqref{parity} and \eqref{parityAK}, can be mapped onto the bicrossproduct basis.

From the ``standard'' definition \eqref{parity} one gets that the action of parity on the bicrossproduct momenta is also standard:
\bea
\mathbb{P} (\bar P_{0})&=& \mathbb{P}\left(  \kappa \ln\left( \frac{P_{0}+P_{4}}{\kappa}\right) \right) =   \bar P_{0}\nonumber\\
\mathbb{P} (\bar P_{i})&=& \mathbb{P}\left(  \kappa \frac{P_{i}}{P_{0}+P_{4}} \right) =  \kappa \frac{ - P_{i}}{P_{0}+P_{4}}  = -\bar P_{i}\,.\nonumber\\
\eea
Given the nonlinear relation between momenta in the two bases this is a non-trivial result.

Mapping the action of  deformed parity \eqref{parityAK} onto the bicrossproduct coordinates one gets an action that is still deformed, but with a  different functional dependence on the momenta:
\bea
\mathbb{P} (\bar P_{0})&=& 
   k \ln\left( \frac{1}{\kappa}\left( P_{0}-\frac{|\vec P|^{2}}{P_{0}+P_{4}}+P_{4} \right) \right) \nonumber\\
&=& \bar P_{0}+\kappa \ln\left(1-\frac{|\vec P|^{2}}{\kappa^{2}} \right)\nonumber\\
\mathbb{P} (\bar P_{i})&=&  \kappa \frac{ -\kappa \frac{P_{i}}{P_{0}+P_{4}}}{P_{0}-\frac{|\vec P|^{2}}{P_{0}+P_{4}}+P_{4}} = -\frac{\bar P_{i} \,e^{-\bar P_{0}/\kappa}}{1-\frac{|\vec{\bar P}|^{2}}{\kappa^{2}}}\,.\nonumber\\
\eea
The fact that the definition of parity on different sets of coordinates is different is not in principle a problem, since we do not expect different bases of $\kappa$-Poincar\'e to be physically equivalent. Nevertheless, this action of parity is again non-idempotent ($\mathbb{P} ^{2}\neq \mathbb{I}$).

While in the embedding basis the algebra is standard and so the definition of the helicity operator is  straightforward, this is no longer the case in the bicrossproduct basis. In particular, we can follow two different strategies to define helicity in the bicrossproduct basis.

The simplest option is to  map the helicity operator obtained in the linearizing basis onto the bicrossproduct coordinates:
\be
\hat h = -\frac{\vec P(\bar P)\cdot \vec J}{|\vec P(\bar P)|} = -\frac{ e^{\bar P_{0}/\kappa}\,\vec {\bar P}\cdot \vec J}{|e^{\bar P_{0}/\kappa}\vec {\bar P}|} = -\frac{ \vec {\bar P}\cdot \vec J}{|\vec {\bar P}|} \,.
\ee
This evidently leads to a definition of helicity that takes the same form in the two bases.

Another option is to define helicity starting from its relation to the Pauli-Lubanski vector:
\be
\hat h = \frac{W^{0}}{P^{0}}.
\ee
In the linearizing basis, because the Puali-Lubanski vector takes the same form as in the standard Poincar\'e algebra, this definition leads to the one used above, eq. \eqref{hclassical}.  In the bicrossproduct basis the symmetry generators satisfy a deformed algebra, so the Pauli-Lubanski vector is modified  \cite{Ruegg:1994bk}, as well as the Casimirs of the algebra. Taking this into account the on-shell helicity operator reads
\be
\hat h =- \frac{\bar W^{0}}{\bar P^{0}}= -\frac{ e^{\bar P_{0}/\kappa}\vec {\bar P}\cdot \vec J}{\bar P^{0}}= \frac{\vec {\bar P}\cdot \vec J}{  (\kappa-|\vec{\bar P}|) \ln\left(1-\frac{|\vec{\bar P}|}{\kappa}\right)} \,,
\ee
where in the last step we used the on-shell condition 
\be
e^{-\bar P_{0}/\kappa}=1-\frac{|\vec{\bar P}|}{\kappa}\,.
\ee 
The eigenvalues of this helicity operator are  energy-dependent, even though for any given energy they can still take two values with opposite signs. This is easily seen by writing 
\be
\hat h = \hat h_{0} \frac{-|\vec {\bar P}|}{  (\kappa-|\vec{\bar P}|) \ln\left(1-\frac{|\vec{\bar P}|}{\kappa}\right)} \,,
\ee
where $ \hat h_{0}$ is the undeformed helicity operator.
Note that the eigenvalues diverge when the spatial momentum reaches the upper bound imposed by the massless on-shell condition, $|\vec{\bar P}|\rightarrow \kappa$.
\\

\subsection{Scale invariance from parity invariance}

The covariant spectrum of fluctuations $P_{\phi}(\vk)$ is defined in terms of the vacuum expectation value of a quantum field as follows
\be
\langle 0| \phi^2 (x) |0 \rangle = \int d\bar{\mu}(\vk) P_{\phi}(\vk)
\ee
where $ d\bar{\mu}(\vk) $ is the covariant on-shell measure which in ordinary QFT has the well known form $ d\bar{\mu}(\vk) = \frac{d \vk}{2 \omega_\vk}$.  In isotropic theories we can factor out the angular dependence in $\langle 0| \phi^2 (x) |0 \rangle$ and write the spectrum and the integration measure just as functions of the norm of the spatial momentum $p$, e.g. in $3+1$-dimensions
\be
\langle 0| \phi^2 (x) |0 \rangle =  4\pi \int d\bar{\mu}(k) P_{\phi}(k)\,
\ee
with $d\bar{\mu}(k) = \bar{\mu}(k) dk$. As shown in \cite{Arzano:2015gda} the dimensionless power spectrum of curvature fluctuations $\mathcal{P}_{\zeta}(k)$ is related to the covariant power spectrum by
\be
\mathcal{P}_{\zeta}(k) \propto G k\, \bar{\mu}(k)\,  P_{\phi}(k). \label{covariant}
\ee
We now denote with $\delta_{\bar{\mu}}$ the Dirac delta associated to the covariant measure i.e. such that 
\be
\int d\bar{\mu}(\vk) \delta_{\bar{\mu}}(\vk) =1\,.
\ee
Given the field operator
\be\label{fieldo}
 \phi (x) =  \int d\bar{\mu}(\vk) (a(\vk)\, e^{-ik x} + a^{\dagger} (\vk) e^{ik x} )
\ee
and the canonical commutator 
\be
[a(\vk), a^{\dagger}(\vk')] =  \delta_{\bar{\mu}}(\vk- \vk')\,, 
\ee
one immediately sees that for {\it any} momentum space measure 
\be
P_{\phi}(k) = 1
\ee
and thus scale invariance of $\mathcal{P}_{\zeta}(k)$ can be fully characterized in terms of the properties of $\bar{\mu}(k)$. We now look at the field operator in the $\kappa$-deformed theory \cite{Arzano:2010jw, Arzano:2009ci}. Two non-trivial ingredients will now enter \eqref{fieldo}. The first is the measure on the curved, de Sitter momentum space \eqref{eq:constraint1} given, in embedding coordinates, by
\be
d\mu(k) = d^4 k \frac{\,\kappa}{ \sqrt{\kappa^2+k_0^2-\vk^2}}\,.
\ee
We immediately notice that for a massless field on-shell such measure reduces to the ordinary flat Lebesgue measure $d^4 k$   and thus adds no further contribution to the ordinary spectrum of fluctuations. The second non-trivial ingredient which enters in the deformed quantum filed is the plane wave whose momentum labels now obey the deformed composition and inverse operations we recalled at the beginning of this Section. In particular we now have that 
\be
\langle 0| \phi^2 (x) |0 \rangle = \int d\bar{\mu}(\vk)\, d\bar{\mu}(\vk')\, \langle \vk | \vk' \rangle\, e^{i (\ominus k \oplus k') x}\,. 
\ee
We have to evaluate the explicit form of 
\be
\langle \vk | \vk' \rangle = \langle 0| [a(\vk), a^{\dagger}(\vk')]  |0 \rangle =   \delta_{\bar{\mu}}(\vk \oplus (\ominus \vk'))\,.
\ee
Since in embedding coordinates on-shell the momentum space measure is the same as in the undeformed case, we have that 
\be
\delta_{\bar{\mu}}(\vk \oplus (\ominus \vk')) = 2\omega_{\vk}\delta^{(3)}(\vk \oplus (\ominus \vk'))
\ee
where $\delta^{(3)}(...)$ denotes the standard delta function. Using the relations
\be
\vk \oplus \vk' =  \frac{1}{\kappa}\vk (k'_{0}+k'_{4})+ \vk'
\ee
and 
\be
\ominus k_0 = -k_0 + \frac{\vk^2}{k_0+k_4}\,,
\ee
the massless on-shell relation $k_0 = |\vk|$ and the fact that $\ominus k_4 = k_4$ we obtain
\be
\delta_{\bar{\mu}}(\vk \oplus (\ominus \vk')) = 2\omega_{\vk}\left(1+ \frac{|\vk|}{\kappa}\right)\delta^{(3)}(\vk - \vk')\,.
\ee
In principle this result is telling us that indeed the non-trivial composition rule is adding an extra momentum dependent term to the inner product
\be
\langle \vk | \vk' \rangle = 2\omega_{\vk}\left(1+ \frac{|\vk|}{\kappa}\right)\delta^{(3)}(\vk - \vk')\,.
\ee
However plugging $\langle \vk | \vk' \rangle$ back in the expression for $\langle 0| \phi^2 (x) |0 \rangle$ one can quickly check that the non-trivial delta function $\delta^{(3)}(\vk \oplus (\ominus \vk'))$ puts to zero the argument of the plane wave in the integral and thus one still has $P_{\phi}(k) = 1$ despite the non-trivial contribution from the addition law. This is consistent with the results found in \cite{Gubitosi:2015osv} using the bicrossproduct basis. The dimensionless power spectrum is then found using eq.\eqref{covariant}, with  standard $\bar\mu(k)=\frac{k^{2}}{(2\pi)^{3}2 k}$:
\be
\mathcal{P}_{\zeta}(k) \propto G\frac{k^{2}}{2 (2\pi)^{3} }. 
\ee
We see that this spectrum is not compatible with invariance under  the deformed parity operator  \eqref{parityAK}:
\be
\PP({\cal P_{\zeta}}(k))={\cal P}(S(k) ) \sim (S(k))^{2} =\frac{\kappa^{2} k^{2}}{(k+\kappa)^{2}}\,,
\ee
as discussed in Section \ref{scaleinvariance}, and thus in this framework the requirement of parity invariance of the spectrum is strictly related with its scale invariance.

\section{Conclusions}
The fact that quantum gravity might lead to parity violations is not new in the literature. This could happen both because the theory acquires a chiral dynamics~\cite{chern,chern2,leecj} either at the classical or quantum level, or because the vacuum itself violates parity~\cite{joao1,joao2}. The observational implications for CMB polarization are also well-known (e.g.~\cite{leecj}). In this paper we went deeper into the problem of parity at the Planck scale, questioning whether the definition of parity itself might have to be modified, with the loss of basic properties often taken for granted. We discovered that enforcing a modified parity would then lead to new effects, potentially parading as apparent violations of conventional parity. 

We considered the problem both in abstract and with reference to concrete models of quantum gravity. The basic problem is that with the loss of a conventional position space and a mirror action there is not reason to enshrine idempotency into the definition of parity. With the potential loss of idempotency we are led to contemplate the loss of the unitarity and the hermiticity of the parity operator. But this could happen, too, with idempotency preserved. In either case it could happen that parity maps a right handed state into a quantum superposition of right and left handed states. 
Enforcing parity would then, itself, lead to apparent violations of parity. It could also be that parity acts non-trivially upon the momentum space, so that one obtains more information than the usual constraints already obtained from isotropy for scalar fluctuations. Indeed we find that a  close relation between parity and scale-invariance emerges in this context. Could, then,  the observed departures from exact scale-invariance be a manifestation of parity asymmetry at the Planck scale?

We then studied the problem with reference to one concrete model for exploring quantum space-time: the $\kappa$-Poincar\'e approach. In this case we showed explicitly how the lost idempotency of the parity transformation is realized in this framework for Planck-scale kinematics. A peculiar feature of the model concerns the helicity operator which, in the presence of deformed parity, maps left and right states into each other changing the value of their spatial momenta. This behaviour leads to an interesting connection between parity invariance and scale invariance of quantum fluctuations which we worked out explicitly.

It would be interesting to explore the general framework  presented at the start of this paper for other theories. An example is rainbow gravity~\cite{rainbow}: the idea that the space-time metric runs with the energy because position space and momentum space have become entangled. How would parity be affected by such a set up? We defer to further work an investigation of the matter for this and other theories of the Planck scale. 

\section{Acknowledgements}

We thank Giovanni Amelino-Camelia and Jerzy Kowalski-Glikman for discussions related to this paper.  GG and JM acknowledge partial support from the John Templeton Foundation. JM was also supported by  an STFC consolidated  grant.


\end{document}